\documentclass[pmlr]{jmlr}

\usepackage{longtable}

\usepackage{booktabs}
\usepackage[load-configurations=version-1]{siunitx} 

\usepackage{hyperref}       
\usepackage{url}            
\usepackage{booktabs}       
\usepackage{amsfonts}       
\usepackage{nicefrac}       
\usepackage{microtype}      
\usepackage{todonotes}
\usepackage{adjustbox}
\usepackage{array, multirow}
\usepackage{makecell}
\usepackage[normalem]{ulem}
\usepackage{arydshln}
\usepackage{comment}
\usepackage{floatrow}
\usepackage{bm}

\usepackage{graphicx}
\usepackage{color,soul}
\usepackage{graphicx}
\usepackage{caption}
\usepackage{floatrow}
\usepackage{adjustbox}
\usepackage{natbib}
\usepackage{booktabs}

\theorembodyfont{\upshape}
\theoremheaderfont{\scshape}
\theorempostheader{:}
\theoremsep{\newline}

\title[Cross-Language Aphasia Detection]{Cross-Language Aphasia Detection using Optimal Transport Domain Adaptation}

 \author{\Name{Aparna Balagopalan} \Email{aparna@winterlightlabs.com}\\
 \addr Winterlight Labs 
 \AND
 \Name{Jekaterina Novikova} \Email{jekaterina@winterlightlabs.com}\\
 \addr Winterlight Labs
 \AND
 \Name{Matthew B. A. McDermott} \Email{mmd@mit.edu}\\
 \addr Massachusetts Institute of Technology
 \AND
 \Name{Bret Nestor} \Email{bretnestor@cs.toronto.edu}\\
 \addr University of Toronto, Vector Institute
 \AND
 \Name{Tristan Naumann} \Email{tristan@microsoft.com}\\
 \addr Microsoft Research
 \AND
 \Name{Marzyeh Ghassemi} \Email{marzyeh@cs.toronto.edu}\\
 \addr University of Toronto, Vector Institute
 }

\begin{document}

\maketitle

\begin{abstract}
Multi-language speech datasets are scarce and often have small sample sizes in the medical domain. 
  Robust transfer of linguistic features across languages could improve rates of early diagnosis and therapy for speakers of low-resource languages when detecting health conditions from speech.  
  We utilize out-of-domain, unpaired,  single-speaker, healthy speech data for training multiple Optimal Transport (OT) domain adaptation systems. 
  We learn mappings from other languages to English 
  and detect aphasia from linguistic characteristics of speech, and show that OT domain adaptation improves aphasia detection over unilingual
  baselines for French (6\% increased F1) and Mandarin (5\% increased F1).
  Further, we show that adding aphasic data to the domain adaptation system 
  significantly increases
  performance for both French and Mandarin, increasing the F1 scores further (10\% and 8\% increase in F1 scores for French and Mandarin, respectively, over unilingual baselines).
\end{abstract}

\section{Introduction}
\label{sec:intro}

\label{aba:sec1}
Aphasia is a form of language impairment that affects speech production and/or comprehension. 
It occurs due to brain injury, most commonly from a stroke, and affects up to 2 million people in the US alone~\cite{aphasiafacts}. Evaluation of speech is an important part of diagnosing aphasia and identifying sub-types. 
Aphasic speech exhibits several common patterns; e.g., omitting short words (``a", ``is" ), using made-up words, etc. Prior work has shown that it is possible to detect aphasia with machine learning (ML) from patterns of linguistic features in spontaneous speech~\cite{fraser2014automated}, but a vast majority of research is restricted to a single language
~\cite{fraser2014automated,le2017automatic, qin2018automatic}.

Cross-linguistic studies for aphasia detection and screening~\cite{bates1991cross, soroli2012linguistic, kristen2014crosslinguistic} are important for translating developments made in resource-rich languages (e.g., English) to other languages. Cross-lingual aphasia detection is a hard task, mainly because there is little prior research indicating what features of language affected by impairment are transferable and due to the small size of datasets in the field (typically between 100-500 subjects).

Existing research on cross-language translation of text, word embeddings and audio has demonstrated the value in using large amounts of paired and unpaired data (with dataset size varying from 5k to 1.5M) by imposing 
constraints such as cycle-consistency or incorporating domain-knowledge~\cite{xu2018unpaired, yang2018unsupervised, conneau2017word,jia2019direct,weng2018mapping,Weng:2019:UCL:3292500.3330710}. These allow translation of representations across languages for a variety of tasks, even with \emph{no aligned data} between languages, which is of interest for our prediction task.

In this work, we study cross-linguistic transfer of aphasia detection models trained on English speech from a multi-lingual dataset of healthy and aphasic speech, AphasiaBank~\cite{macwhinney2011aphasiabank}. 
We featurize our speech via the proportions of 8 standard linguistic Part-of-Speech (POS) tags from speech transcripts of each language using the StanfordNLP library~\cite{qi2018universal}, motivated by prior work on automatic aphasia detection with text features~\cite{fraser2014automated}. 
We adapt features from different languages to English using a state-of-the-art technique for domain adaptation, Optimal Transport (OT), with a large \emph{single-speaker, unpaired multilingual dataset} of TED talk transcripts~\cite{tiedemann2012parallel}. OT is a method of domain adaptation where the cost of moving samples from source to target probability distributions is minimized (Sec.~\ref{sec:optimal_transport}
). In our case, probability distributions are the distributions of linguistic features in each language.

We benchmark performance of models~\cite{pedregosa2011scikit,flamary2017pot} trained on English AphasiaBank and tested on French and Mandarin AphasiaBank using different levels of capacity, including support vector machines (SVM), Random Forests (RF), and fully-connected neural networks (NN). We compare unilingual models, which suffer due to the limited size of the French and Mandarin datasets, to two kinds of domain adaption: a baseline multi-lingual autoencoder, and OT based domain adaption. While a multi-lingual autoencoder works well for the similar languages of French and English, it performs poorly for Mandarin. In contrast, OT-based adaption systems perform well across both languages. We additionally study the impact of various settings on the OT-domain adaptation system, such as inclusion of paired data, addition of in-domain data and speech-accent, against both unilingual baselines and an autoencoder-based domain adaption model. We show that OT-domain adaptation with Earth Movers Distance (EMD) and entropic regularization achieves 10\% and 8\% increase in F1 scores over the unilingual baselines in French and Mandarin respectively. From our analysis, we identify that large datasets, which could be unpaired across languages, but also include some samples of aphasic speech are essential for higher detection rates. 
 
%
\paragraph{Healthcare Relevance:} Medical speech datasets for many languages are small and few in number. As a result, most of the prior work on computational methods for detecting signs of aphasia focuses on ML-models developed for a single language~\cite{fraser2014automated,le2017automatic, qin2018automatic}, or cross-language feature patterns for a single feature~\cite{bates1991cross}. In this work, we study both feature transfer across similar 
and dissimilar 
languages when participant speech is represented by multiple linguistic features, as well as cross-linguistic transfer of ML-models for detecting aphasia from linguistic features for translating developments made in English to other languages.
\paragraph{Technical Sophistication:} 
We benchmark the utility of unpaired speech datasets with OT to demonstrate a use of cross-language domain adaptation to
account for sparsely labeled languages in a clinical speech task. We perform rigorous ablation studies to investigate the effect of paired data (inclusion of paired data does not increase F1-scores significantly
), the effect of aphasic samples in domain adaptation (improves F1-scores significantly over unilingual baselines and domain adaptation with unpaired data, performs on the same or significantly higher level than multi-lingual baselines) and the effect of diversity of speech samples in terms of accents (more diverse data for OT is better).
While we only compare OT-domain adaptation with a multi-lingual autoencoder, and do not compare against other, more recent techniques, such as adversarial domain adaptation~\cite{tzeng2017adversarial},
we demonstrate that a state-of-the-art technique for domain adaptation, Optimal Transport, can be used to improve aphasia detection in a cross-language evaluation setting.

\section{Background \& Related Work}

Existing studies have considered machine learning (ML) based approaches to aphasia detection~\cite{fraser2014automated, le2017automatic}, but a summary of previous work reveals these have been largely restricted to single language settings (Tab.~\ref{tab:prevwork}). While \cite{10.3389/conf.fnhum.2018.228.00075} considers multiple languages, only non-speech-related features were used and there was no model transfer across languages. To the best of our knowledge, there is no prior work where cross-language transfer has been used in the detection of aphasia.

\begin{table*}[h]
\centering
\caption{Summary of aphasia analysis in various languages, including usage of any external corpus in addition to the $N$ reported (indicative of training/test size), feature modalities and performance \label{tab:prevwork}}
{
\begin{adjustbox}{max width=1\linewidth}
\begin{tabular}[t]{lccccc}
\hline
Methods &Language(s)&Features&Subjects/Samples (N,M) &External corpus (Y/N) &Performance\\
\hline
\hline
~\cite{fraser2014automated} & English& Linguistic & 26, 26 &N&1.00 (Accuracy)\\
~\cite{10.3389/conf.fnhum.2018.228.00075}& German, Italian and Greek&  Linguistic, cognitive performance indices &26, 104 & N & 0.79 (AUC) \\
~\cite{law2018analysis} & Cantonese & Lexical and semantic & 65, 65 & N & -\\
~\cite{qin2018automatic} & Cantonese & Text and acoustic & 82, 328 & N & 0.90 (F1-score)\\
~\cite{ishkhanyan2017grammatical} & French & Lexical & 15, 45 & N & -\\
Ours & French & Text & 24, 24 & Y & 0.87 (F1-score) \\
Ours & Mandarin & Text & 60, 55 & Y & 0.69 (F1-score) \\
\end{tabular}

\end{adjustbox}
}
\end{table*}


\subsection{Domain Adaptation}
\paragraph{Optimal Transport for Domain Adaptation:} Diverse methods have been explored for domain adaption. Methods involving adversarial loss have been developed for multi-lingual word embeddings and language translation  \cite{conneau2017word, xu2018unpaired}; some of them specialized to clinical machine translation \cite{weng2018mapping}. We use Optimal Transport, an embedding-based method of domain adaptation (Sec.~\ref{sec:optimal_transport}). Variants of the base OT approaches have been proposed for a variety of NLP tasks in prior work \cite{chen2019improving,courty2017optimal}, including aligning representations across domains in an unsupervised manner \cite{bhushan2018deepjdot}. OT-based sequence-to-sequence learning techniques have outperformed strong baselines in machine translation and abstractive summarization~\cite{chen2019improving}, while modifications to the base OT algorithms have set new benchmarks for unsupervised word translation~\cite{alvarez2019towards}.

\textbf{Cross-linguistic Adaptation:} Recent work in dementia detection, rather than aphasia, used paired samples from the OpenSubtitles~\cite{lison2016opensubtitles2016} dataset to train a regression model between independently engineered features from Mandarin and English transcripts~\cite{li2019detecting}. In contrast, we use unpaired data and learn mappings between \emph{distributions} of the same linguistic features between different source languages and English. We utilize unpaired datasets in our study since 
this approach is more general and more useful when paired datasets are not available between a resource-rich 
and other languages.
OT overcomes the requirement of paired data by aligning probability distribution functions of the linguistic features, rather than the features themselves.

\section{Data Sources and Pre-processing}
In this section, we provide details regarding all our data sources and text preprocessing steps.
\subsection{AphasiaBank}
All datasets of speakers of English, French and Mandarin are obtained from AphasiaBank~\footnote{\url{https://aphasia.talkbank.org/}}~\cite{macwhinney2011aphasiabank}. The aphasic speakers have various subtypes of aphasia - broca, wernicke, anomic, etc. (See App.~\ref{app:aphasia}
) All participants perform multiple speech-based tasks, such as describing pictures
, story-telling, free speech and discourse. We combine all tasks to a single transcript in our analysis.
Detailed statistics for each language are in Table ~\ref{tab:aphasiabank}. 
All samples are manually transcribed, following the CHAT protocol ~\cite{ratner1993brian}. We classify speech samples to two classes - healthy and aphasic, where aphasic constitutes all sub-types mentioned above, using extracted linguistic features.

\subsection{TED Talks}
We use a large dataset of TED talks with multi-lingual transcripts \cite{tiedemann2012parallel} to train our domain adaption systems. In total, there are recordings available for 1178 talks, with various speaker accents and styles. We use transcripts from Mandarin, French and English languages. 
So that our domain adaption system is not biased by seeing paired data, which is not present in our aphasia classification task, we ensure there is no overlap between speech transcripts of English and French/Mandarin by dividing the talks into two sets and ensuring that the English transcripts for training the domain adaptation models are obtained from the first set, while those for French/Mandarin are obtained from the second set.\footnote{We also performed experiments to validate that this choice does not significantly affect results, finding that using either fully paired or fully unpaired (as described here) data yields statistically insignificantly different results.}
Additionally, similar to the methodology of~\cite{li2019detecting}, we attempt to create a larger dataset by dividing each narration into segments by considering 25 consecutive utterances as one segment. We choose 25 because we observed that the features stabilize with this number of utterances (see App.~\ref{app:length} for details).

\begin{table}[h]
\caption{Number of samples from AphasiaBank and the TED Talks corpus. Number of participants indicated in parentheses.\label{tab:aphasiabank}
}
{
\begin{adjustbox}{max width=0.5\linewidth}

\begin{tabular}[t]{lcccc}
\hline
Corpus & Language &Healthy samples &Aphasic samples\\
\hline
\hline
AphasiaBank & English  & 246  (192) &  428 (301) \\
AphasiaBank & French   & 13   (13)  & 11 (11)\\
AphasiaBank & Mandarin & 42   (40)  & 18 (15)\\
TED Talks   & English  & 2875 (589) & - \\
TED Talks   & French   & 2976 (589) & - \\
TED Talks   & Mandarin & 2742 (589) & - \\
\end{tabular}
\end{adjustbox}
}
\end{table}

\subsection{Transcript Pre-processing and Feature Extraction}
\label{sec:feature_extraction}
The transcripts provided in AphasiaBank consist of transcribed speech following the CHAT protocol~\cite{macwhinney2011aphasiabank}. Hence, there are several annotations such as repetitions, markers for incorrect word usage etc. To extract features, an important pre-processing step is to remove these various additional annotations. We utilize the \emph{pylangacq}~\cite{lee-et-al-pylangacq:2016} library for this step, due to its capabilities of handling CHAT transcripts.
Additional pre-processing steps include stripping the various utterances of punctuations before POS-tagging. We  extract  the  proportion  of  8 
POS\footnote{\url{https://universaldependencies.org/u/pos/}} -- nouns,  verbs,  subordinating conjunctions, adjectives, adverbs, coordinating conjunctions, determiners and pronouns -- over the whole transcript of speech. Though aphasic  speakers perform one additional speech task (where they provide details regarding their stroke) more than control speakers,  these 8 features are agnostic to total length and content of transcripts, and rely more on the sentence complexity. 
These simple features  are used because they are general and have been identified to be important in prior work~\cite{fraser2013using, li2019detecting, law2013production} across languages.  These features are extracted from all languages in AphasiaBank.

To analyse the variance in features across languages, we study if they differ significantly between healthy and aphasic speakers across languages (Tab.~\ref{tab:pvalues} in the App.). We observe that every feature varies significantly between healthy and aphasic speakers of English and Mandarin. We anticipate, hence, that raw, non-adapted 
cross-language transfer of models trained on English speech to Mandarin would lead to low performance.

\section{Methods}

\begin{figure}
  \includegraphics[height=2.7in]{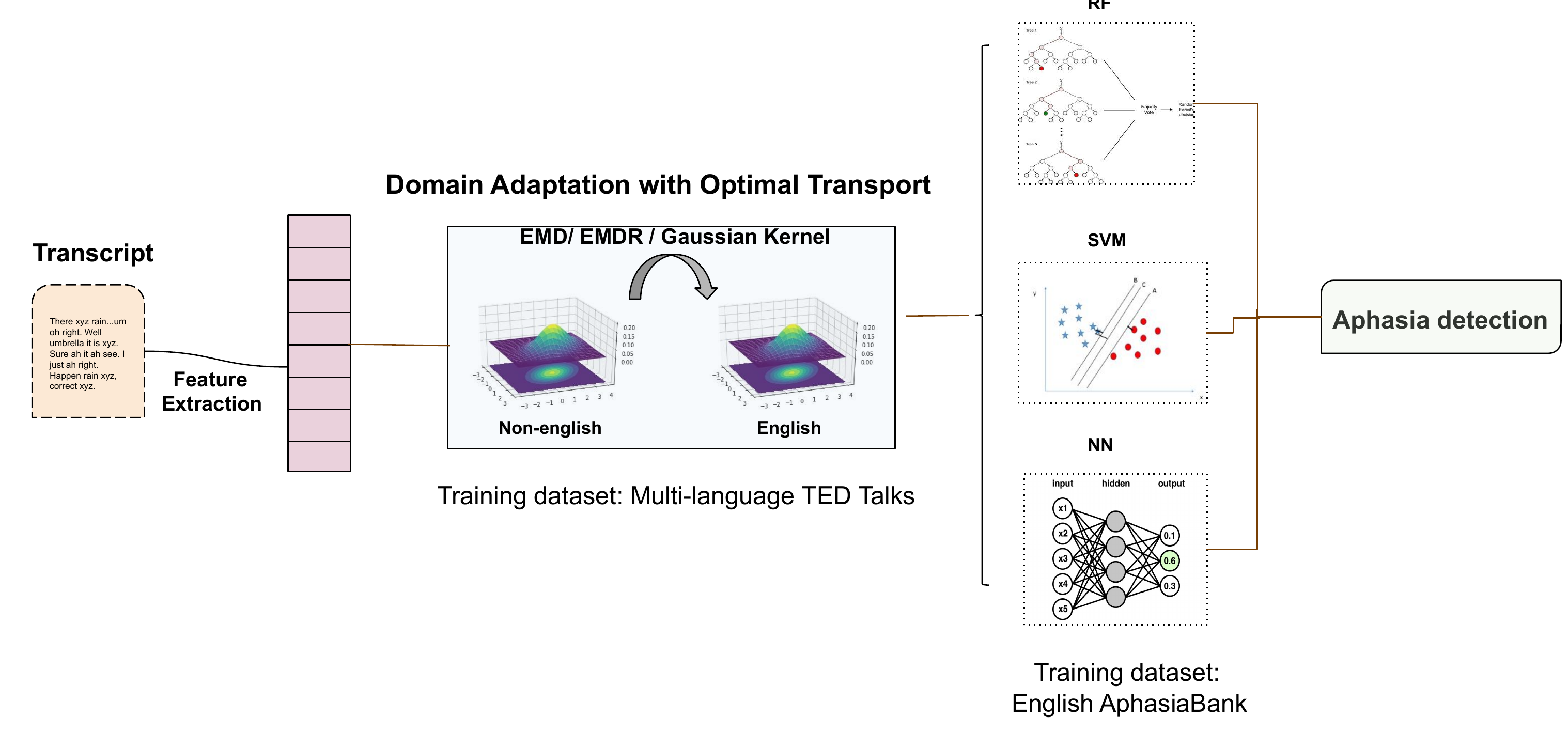}
  \caption{Pipeline for processing a speech transcript from a non-English language. Features are extracted as detailed in Sec.~\ref{sec:feature_extraction}, cross-lingual representations are obtained with multiple Optimal Transport algorithms, and aphasia detection classification performed with different ML models.}
  \label{fig:pipeline}
\end{figure}

We describe the domain adaptation system, which uses Optimal Transport. Overall pipeline in Fig.~\ref{fig:pipeline}.

\subsection{Cross-linguistic Representation Learning with Optimal Transport}
\label{sec:optimal_transport}
Optimal transport (OT) consists of finding the best transport strategy from one probability distribution function (PDF) to another. This is done by minimizing the total cost of transporting a sample from the source to that in the target. Thus, there needs to be a metric to quantify the different distances between samples in the two probability distributions, as well as solvers to solve the optimization problem of minimizing the total cost of transport, where cost is related to distance between source and target. We use optimal transport for domain adaptation here because we extract the same features across languages, though their distributions (in terms of feature values, e.g. proportion of nouns) vary from one language to another. 

We use three solvers and distance functions between PDFs based on optimal transport:
\begin{description}
    \item[Earth Movers Distance OT (EMD)\label{method:emd}] EMD or Wasserstein distance between the two distributions is minimized using an optimal transport Network Flow \cite{bonneel2011displacement}.

    \item[Gaussian Optimal Transport Mapping (Gaussian kernel)] The Earth Movers Distance (EMD) or Wasserstein distance between the two distributions is minimized same as in \ref{method:emd}. However, the transport map is approximated with a gaussian kernelized mapping to obtain smoother transport maps~\cite{perrot2016mapping}.

    \item[Entropic Regularization OT solver (EMD-R)] Optimal transportation problem with EMD regularized by an entropic term, turning the linear program into a strictly convex problem that can be solved with the Sinkhorn-Knopp matrix scaling algorithm \cite{sinkhorn1967concerning}. Linear solver proposed by ~\cite{cuturi2013sinkhorn} is used.
\end{description}




We employ open-source implementations of these algorithms~\cite{flamary2017pot}. We will refer to each of these OT algorithms as \emph{EMD, Gaussian kernel, EMD-R} respectively as above. Detailed hyperparameter settings for early stopping tolerance, regularization terms (in EMD-R) are are in Appendix~\ref{app:ot_hyperparams}. 

OT mappings from each source language (French/Mandarin) to English are learned for each algorithm, trained on language pairs (English-French/English-Mandarin) from the TED Talks dataset.


 

\section{Experiments}
\label{sec:classification}

We consider the classification of speech as aphasic or healthy from speech transcripts (as featurized via POS proportions), and we are primarily interested in whether performance can be improved for low-resource languages.
%
This classification task is performed across several baseline settings, including a unilingual task, direct feature transfer, and an autoencoder-mediated multilingual encoding, as well as various domain adaptation settings (Sec.~\ref{sec:training_regimes_ot}), where features are mapped from low resource languages to English using Optimal Transport. A suite of ML models, including SVM, RF and NN are used for this task (see Sec.~\ref{sec:experiments} for hyperparameters).

We evaluate task performance primarily using macro-averaged F1 scores, a measure that is known to be robust to class imbalance.
We also report AUROC scores, since it is often used as a general measure of performance irrespective of any particular threshold or operating point~\cite{richardson2006markov,liu2007comparing}.

Due to the lack of baselines on the multilingual AphasiaBank dataset in prior work, we establish our own unlingual and multilingual baselines, detailed in the section below.

\subsection{Baseline Domain Adaptation Systems}
\label{sec:baselines}

\paragraph{Unilingual Training:} Unilingual baselines for each language using 10-fold cross-validation (CV), stratified by subject so that each subject’s samples do not occur in both training and testing sets in each fold. We estimate that this would be a lower bound on performance for French and Mandarin AphasiaBank, since it is likely, given the small size of the dataset, that models would underfit and have low generalizable performance across subjects.

\paragraph{Feature Transfer from English with raw, non-adapted features:} We also identify transfer baselines, wherein models trained on English AphasiaBank are evaluated on other languages with no fine-tuning.  We hypothesize that this baseline would be more performant than the uni-lingual baseline, at least amongst the more-similar Romance languages of French and English, since it utilizes the comparatively larger dataset of English AphasiaBank for training. 

\paragraph{Multilanguage Embedding with an Autoencoder:}
A common representation is obtained for all three languages by encoding the linguistic features using a high capacity autoencoder. This autoencoder, trained on English, French and Mandarin TED Talks datasets (unpaired), maps linguistic features extracted  from  multilingual transcripts into a shared latent space. The autoencoder consists  of  4  hidden layers  (2 hidden layers in encoder and decoder respectively) with 5, 3, 3 and 5 units each for the following experiment.  Hyperparameters are set using a 90-10 train-dev split of samples from each language. All ML classifiers then are trained on the encoded versions of English AphasiaBank, and tested on encoded versions of French and Mandarin AphasiaBank. Comparison of other training regimes to this baseline would determine if learning a shared representation across multiple languages is better than OT.




\subsection{Training Regimes for OT Adapted-transfer}
\label{sec:training_regimes_ot}

We evaluate two OT-training regimes for both French and Mandarin, each tested across all varieites of OT (e.g., OT-EMD, OT-Gaussian, and OT-EMD-R).

\textbf{Feature Transfer from English with OT domain adaptation, with TED Talks for OT}:
Models trained on English AphasiaBank evaluated on other languages with OT adaptation (EMD, EMD-R, Gaussian mapping) with no fine-tuning. OT models here are trained only on the multi-language unpaired TED Talks dataset, i.e, \emph{with no aphasic data}. 

\textbf{Feature Transfer from English with OT domain adaptation, with TED Talks and AphasiaBank for OT}:
Models trained on English AphasiaBank evaluated on other languages with OT adaptation (with EMD, EMD-R, Gaussian mapping) with no fine-tuning. The OT models here are trained on the multi-language TED Talks dataset, and multi-language AphasiaBank i.e, \emph{with aphasic data}. We ensure that there is no overlap in the proportion of AphasiaBank used for learning OT mappings and for evaluating the classifiers by employing 2-fold cross-validation where one fold is included in the training set for OT and another for evaluation. 
Since OT involves source and target domain probability estimation, we hypothesize that adding in-domain data, particularly that of speech-impaired participants, would improve results significantly. 

\subsection{Impact of Including Diverse Speech Samples in OT Training}
Since literature shows that accents can have a significant effect on POS features~\cite{runnqvist2013disadvantage}, we hypothesize that there is an observable effect of diversity in terms of accents in the OT training set. To study this effect, we manually annotate accents for the English TED Talks dataset (details in App.~\ref{app:accent_annot}) as `North American'(NA) accent or `other' accent. In total in the TED Talks English set, there are 373 NA accented and 215 `other' accented talks. We study the impact of increasing the diversity of accents used for training OT algorithms, keeping the size of the dataset constant (see Tab.~\ref{tab:accent_ot}).

\subsection{Hyperparameter Settings}
\label{sec:experiments}
Hyperparameters for classification models are tuned using grid search with 10-fold cross validation on the training set (English AphasiaBank) across all settings.
We use an SVM (RBF kernel with regularization parameter $C= 0.1$ and $\gamma= 0.001$), Random Forest (RF; 200 decision trees with a maximum depth of 2), and Neural Network (NN; 2 hidden layers of 100 units each) classifiers for the cross-linguistic classification task\cite{pedregosa2011scikit}. Since the training set is highly imbalanced (see Tab.~\ref{tab:aphasiabank}), the minority class is oversampled synthetically with SMOTE~\cite{chawla2002smote} with $k=3$.
Prior to oversampling, the training set is normalized and scaled using the median and interquartile range, a common mechanism to center and scale-normalize data which is robust to outliers~\cite{pedregosa2011scikit}. The same median and interquartile (obtained from the training set) bounds are used to scale the evaluation set in each case.

All ML classifiers are trained completely only on English AphasiaBank, while the OT models are trained on unpaired samples across English and another language (French or Mandarin) from the TED talks corpus~\cite{tiedemann2012parallel}.

\section{Results}
\label{sec:results}
In Tab.~\ref{tab:transcript_performance}, we compare the performance of various OT-algorithms to their respective baselines for cross-language representation learning for the aphasia detection prediction task. 

\paragraph{Baselines}
We see that baseline performance varies significantly between languages. For French, using a multilingual encoding or direct feature transfer largely offers significant improvements over unilingual training, yielding a maximal lift of 15 for RF mean F1, and achieving maximal overall classifier performance using multilingual encoding with an SVM model. In general for French, the multilingual encoding outperforms the feature transfer baseline, but both improve on unilingual results.

For Mandarin, these results are very different; here, either baseline approach to adaption hurts overall performance as compared to a unilingual baseline, often yielding solutions for which the model will predict just a single class exclusively.


\begin{table*}[ht!]
\caption{F1 macro and AUROC mean and standard deviation scores across languages for different model settings in OT training, averaged across multiple runs. Note that zero standard deviations occur when a single class is predicted or when standard deviation $<0.01$. The standard deviations are an artefact of the small sample sizes in the evaluation set (24 and 60 for French and Mandarin respectively), as seen in prior literature~\cite{fraser2013using}. Highest F1 scores are shown in bold for each language and classifier. Overall, the highest mean F1 scores are obtained with OT adaptation with the EMD variants, with aphasic samples included in OT training, along with the multilingual TED Talks dataset. \label{tab:transcript_performance}. A study on the effect of data size on uni-lingual performance in English is in App.~\ref{app:data_impact_unilingual} for comparison.}
{
\begin{adjustbox}{max width=1\textwidth}
\begin{tabular}{c|l|c|c|c}
\hline
  \multicolumn{1}{l}{\textbf{Language}} &
  \multicolumn{1}{c}{\textbf{Method}} &
  \multicolumn{1}{c}{\textbf{SVM}}& 
  \multicolumn{1}{c}{\textbf{RF}}&
  \multicolumn{1}{c}{\textbf{NN}} \\ 
  \hline
& & F1 \hspace{0.4cm} AUROC&F1 \hspace{0.4cm} AUROC &F1 \hspace{0.4cm} AUROC \\ 
  
  \hline \hline
  \multirow{2}{*}{\makecell{French}} 
  & Unilingual Baseline &  $74.00\pm 0.00$\hspace{0.4cm}$80.00\pm 0.00$ & $64.00\pm 5.44$\hspace{0.4cm}$72.50\pm 4.08$& $76.67 \pm 0.00\hspace{0.4cm}79.17\pm 2.30$ \\
  & Mutlilingual Encoding & $ \mathbf{85.58\pm 3.79}$\hspace{0.4cm}$\mathbf{85.52 \pm 4.07}$  & $ 79.57 \pm5.25 $\hspace{0.4cm}$79.44 \pm 4.90$ & $81.53\pm 4.70$\hspace{0.4cm}$81.96\pm 4.72$  \\
  & Feature Transfer &  $79.13\pm 0.00$\hspace{0.4cm}$78.61\pm 0.00 $ & $77.23 \pm 0.00$\hspace{0.4cm}$77.27\pm 0.00$& $52.93 \pm 5.04$\hspace{0.4cm}$58.97\pm 1.81$ \\
  & $OT$-EMD & $79.13\pm 0.00\hspace{0.4cm}79.38\pm 0.00$ & $ 80.49 \pm 1.92\hspace{0.4cm}81.12\pm 2.57$& $65.18\pm 1.94\hspace{0.4cm}64.80\pm 2.14$ \\
  & $OT$-Gaussian & $53.13\pm 0.00\hspace{0.4cm}61.54\pm 0.00$ & $41.89\pm 3.38\hspace{0.4cm}56.41\pm 1.81$& $39.50\pm 0.00\hspace{0.4cm}53.85\pm 0.00$ \\
  & $OT$-EMD-R & $83.22 \pm 0.00\hspace{0.4cm}83.22\pm 0.00$ & $ 81.76\pm 2.07\hspace{0.4cm}81.70\pm 2.14$& $ 73.48 \pm 1.91\hspace{0.4cm}74.94\pm 1.81$ \\
  & $OT$-EMD - with aphasic  & $83.10 \pm 0.00$\hspace{0.4cm}$84.52 \pm 0.00$ & $ 80.26 \pm 2.46$\hspace{0.4cm}$82.14 \pm 2.06$& $78.84\pm 0.00$ \hspace{0.4cm}$80.95 \pm 0.00$\\
  & $OT$-Gaussian - with aphasic & $45.96 \pm 0.00$\hspace{0.4cm}$58.33 \pm 0.00$ & $39.50 \pm 0.00$\hspace{0.4cm}$54.17 \pm 0.00$& $39.50\pm 0.00$\hspace{0.4cm}$54.17 \pm 0.00$ \\
  & $OT$-EMD-R - with aphasic & $\mathbf{87.23 \pm 0.00}$\hspace{0.4cm}$\mathbf{88.09 \pm 0.00}$ & $ \mathbf{83.10\pm 0.00}$\hspace{0.4cm}$\mathbf{84.52 \pm 0.00}$& $\mathbf{81.68 \pm 2.45}$\hspace{0.4cm}$\mathbf{83.33 \pm 2.07}$ \\
  \hline
   \multirow{2}{*}{\makecell{Mandarin}} 
  & Unilingual Baseline & $60.47\pm 0.00\hspace{0.4cm}57.92\pm 0.00 $ & $57.78\pm 0.54\hspace{0.4cm}60.08\pm 1.02$& $57.66 \pm 2.91\hspace{0.4cm}57.19\pm 1.22$ \\
  & Mutlilingual Encoding & $23.08 \pm 0.00$\hspace{0.4cm}$ 50.00\pm 0.00$  & $  30.92\pm13.38 $\hspace{0.4cm}$51.19 \pm 4.12$& $23.08\pm 0.00$\hspace{0.4cm}$50.00 \pm 0.00$ \\
  & Feature Transfer &  $23.08\pm 0.00\hspace{0.4cm}50.00\pm 0.00 $ & $23.08\pm 0.00\hspace{0.4cm}50.00\pm 0.00$& $23.08\pm 0.00\hspace{0.4cm}50.00\pm 0.00$ \\
  & $OT$-EMD & $ 63.28\pm 0.00\hspace{0.4cm}67.06\pm 0.00$ & $ 55.41\pm 0.06\hspace{0.4cm}57.01\pm 0.67$& $53.79\pm 1.32\hspace{0.4cm}59.92\pm 2.34$ \\
  & $OT$-Gaussian & $31.80\pm 0.00\hspace{0.4cm}51.98\pm 0.00$ & $30.26 \pm 1.09\hspace{0.4cm}51.19\pm 0.56$& $27.11 \pm 0.00\hspace{0.4cm}49.60\pm 0.00$ \\
  & $OT$-EMD-R & $\mathbf{66.25 \pm 0.00}\hspace{0.4cm}67.46\pm 0.00$ & $ 54.43 \pm 1.85\hspace{0.4cm}58.33\pm 3.29$& $56.44 \pm 2.20\hspace{0.4cm}61.11\pm 0.85$ \\
  & $OT$-EMD - with aphasic & $65.59\pm 0.00$\hspace{0.4cm}$\mathbf{70.57 \pm 0.00}$ & $ \mathbf{69.05\pm 1.34}$\hspace{0.4cm}$\mathbf{68.10 \pm 0.91}$& $55.92\pm 2.84$\hspace{0.4cm}$59.00 \pm 3.18$ \\
  & $OT$-Gaussian - with aphasic & $34.75\pm 0.00$\hspace{0.4cm}$54.32 \pm 0.00$ & $32.92 \pm 0.00$\hspace{0.4cm}$53.18 \pm 0.00$& $26.82 \pm 0.00$\hspace{0.4cm}$49.77 \pm 0.00$ \\
  & $OT$-EMD-R - with aphasic& $59.32 \pm 0.00$\hspace{0.4cm}$59.09 \pm 0.00$ & $ 61.18 \pm 0.83$\hspace{0.4cm}$60.23 \pm 0.99$& $\mathbf{62.57 \pm 0.12}$\hspace{0.4cm}$\mathbf{64.39 \pm 0.06}$\\
  
   \hline
   \multirow{1}{*}{\makecell{English}}
  & Unilingual Baseline & $85.89\pm 0.00\hspace{0.4cm}88.93\pm 0.00 $ & $82.14\pm 0.31\hspace{0.4cm}85.01\pm 0.11$& $88.07 \pm 0.35\hspace{0.4cm}88.49\pm 0.52$\\
\end{tabular}
\end{adjustbox}
}
\end{table*}

\paragraph{OT-Variants}
Among OT-Variants, we see generally stronger performance as compared to the unilingual models and baseline domain adaption systems as well. In all but one case, the best-performing OT- variant for a given model/language yields a statistically significant improvement over the best baseline model according to a paired $t$-test, the notable exception being for the SVM model on French text, which does not achieve statistical significance. In general, EMD variants of OT (including both OT-EMD-R and OT-EMD) tend to perform better than OT-Gaussian, and nearly universally, including aphasic speech samples in the OT model yields significant lifts, yielding best in class mean-F1 of 87.23 for an SVM model over French samples under the OT-EMD-R model, or 69.05 for a RF model over Mandarin text via the OT-EMD model.

\paragraph{Speech Diversity}
We additionally analyze how speech diversity, as measured by frequency of various accents in the speech data, affects the performance of OT-EMD-R domain adaption for SVM models in French and Mandarin. Results for this are shown in Table~\ref{tab:accent_ot}. For both French and Mandarin in this case, we observe that increasing the prevalence of non-North American accents in the domain adaption task improves downstream aphasia/non-aphasia classification performance by several F1 points (yielding a score of $87.48$ for French and $69.19$ for Mandarin). Note that these results are not statistically significantly different than the best results found previously in Tab.~\ref{tab:transcript_performance}.

\begin{table*}[ht!]
\caption{F1 macro scores across languages with OT-EMDR, with varying proportions of data\label{tab:accent_ot}. Highest scores are shown in bold. Note that we don't report standard deviation since it $<$ 0.01 in all cases.}
{
\begin{adjustbox}{max width=0.5\textwidth}
\begin{tabular}{c|l|c|c}
\hline
  \multicolumn{1}{l}{\textbf{Language}} &
  \multicolumn{1}{c}{\textbf{Method}} &
  \multicolumn{1}{c}{\textbf{OT Dataset Size}} &
  \multicolumn{1}{c}{\textbf{SVM}} \\
  \hline
& &   &F1\hspace{0.4cm} AUROC \\ 
  
  \hline \hline
 
  \multirow{2}{*}{\makecell{French}} 
  &$OT$-EMD-R & 286 NA &$83.22\hspace{0.4cm}83.22$\\
  &$OT$-EMD-R &215 NA, 71 not NA &$83.22\hspace{0.4cm}83.22$\\
  &$OT$-EMD-R &143 NA, 143 not NA &$83.22\hspace{0.4cm}83.22$\\
  &$OT$-EMD-R &71 NA, 215 not NA &$\mathbf{87.48\hspace{0.4cm}87.76}$\\
  \hline
   \multirow{2}{*}{\makecell{Mandarin}} 
  &$OT$-EMD-R & 286 NA&$66.25\hspace{0.4cm}67.46$\\
  &$OT$-EMD-R  & 215 NA, 71 not NA&$62.50\hspace{0.4cm}63.49$\\
  &$OT$-EMD-R  & 143 NA, 143 not NA&$68.51\hspace{0.4cm}70.24$\\
  &$OT$-EMD-R &71 NA, 215 not NA &$\mathbf{69.19\hspace{0.4cm}71.83}$\\
\end{tabular}
\end{adjustbox}
}
\end{table*}

\section{Discussion}
\label{sec:discussion}

\paragraph{Direct Feature Transfer Only Relevant in Similar Languages}
\label{sec:direct_transfer}
In Tab.~\ref{tab:transcript_performance}, we observe that direct feature transfer (i.e., the ``Feature Transfer'' row) achieves good performance for the English to French domain adaption task, but not for the English to Mandarin adaption task. This makes sense as English and French have relatively similar grammatical patterns~\cite{roberts2012verbs} (e.g., subject, verb, object ordering) whereas Mandarin and English have a number of significant differences, including, e.g., reduplication, where a syllable or word is repeated to produce a modified meaning, in Mandarin \cite{li1989mandarin}.

Relatedly, the multilingual encoding approach likewise yields good performance only for French. Here, we again note that French and English are relatively similar languages, compared to English and Mandarin. Thus, our multilingual encoder may be much more able to jointly encode English and French than it could English and Mandarin.

\paragraph{Inclusion of Aphasic Samples is Highly Impactful on OT-performance}
We observe (Tab.~\ref{tab:transcript_performance}), that the highest mean F1-score for cross-language classification on the evaluation set increases to $87.23$ (\emph{OT-EMD-R} with SVM) for French and $69.04$ (\emph{OT-EMD} with SVM) for Mandarin from $83.22$ and $66.25$ respectively (both significant increases, with $p<0.001$ and $p=0.015$ respectively) with the addition of aphasic samples in the training set for OT adaptation. This demonstrates that including aphasic samples has a strong positive effect on OT- based domain adaption. Note that similar results of performance improvement due to addition of in-domain, speech-impaired data have also been observed for multi-lingual topic modelling from speech in prior literature~\cite{fraser2019multilingual}.
\label{sec:in_domaindata}


\paragraph{Diverse Speech Samples in Representation Improve Performance}
\label{sec:accent}
As stated in Section~\ref{sec:results}, we find that increasing the diversity of our OT dataset (as measured through accent distribution) has a positive effect on downstream transfer. This resonates with prior findings, which have shown that accents can have a significant effect on POS features~\cite{runnqvist2013disadvantage}.
\section{Conclusions and Future Work}

A limitation of our current work is that it focuses mainly on a single method of domain adaptation, Optimal Transport. Additionally, the feature set is limited to only include text-based features and hence, results are dependant on the features extracted, and change in feature space might have a significant effect on the relative performance of domain adaptation and multilingual representation learning.
In future work, we will empirically compare OT domain adaptation strategy with other techniques, such as adversarial domain adaptation~\cite{tzeng2017adversarial} for the  aphasia detection task. Furthermore, we plan to study the effect of different featurizations (such as voice acoustics and inclusion of more linguistic features.) on overall performance with the current setup. 

Availability of datasets of an appropriate quality and size is essential in the ML for Healthcare domain, due to the high cost of errors, as well as to ensure fair decisions for all individuals~\cite{rajkomar2018ensuring}. Various solutions have been proposed previously for mitigating the problem of data availability, including creating novel sources of data, developing data-efficient algorithms, and employing domain adaptation from low-resource to resource-rich domains~\cite{li2019detecting,bull2018active}. However, the importance of standard, diverse, in-domain medical speech datasets is underscored by our observations made in Sec.~\ref{sec:discussion}. 


In summary, we show that POS features extracted from speech transcripts from different languages can be mapped to English to aid in clinical speech classification task. 
We find that the OT strategy is successful in domain adaptation,  with associated increase in classification performance for French and Mandarin over unilingual baselines. In comparison to a multilingual baseline with a high-capacity autoencoder, OT algorithms work on par for similar, and significantly better for dissimilar languages. Our results suggest that domain adaption strategies, in particular OT-based domain adaption, can help enable strong predictive models for aphasia detection in low-resource languages

\acks{Dr. Marzyeh Ghassemi is funded in part by Microsoft Research, a CIFAR AI Chair at the Vector Institute, a Canada Research Council Chair, and an NSERC Discovery Grant.}

\bibliography{jmlr-camera-ready}

\clearpage
\appendix

\section{Aphasia sub-types}
\label{app:aphasia}
\begin{itemize}
    \item Broca aphasia or non-fluent aphasia: Individuals with Broca’s aphasia have trouble speaking fluently but their comprehension can be relatively preserved.
    \item Wernicke's aphasia or fluent aphasia: In this form of aphasia the ability to grasp the meaning of spoken words is chiefly impaired, while the ease of producing connected speech is not much affected.
    \item Anomic aphasia: Individuals with anomic aphasia can understand speech and read well but frequently are unable to obtain words specific to what they wish to talk about -- particularly nouns and verbs.
    \item Transcortical aphasia: Individuals with this type of aphasia have reduced speech output, typically due to a stroke.
    \item Conduction aphasia: Individuals with can comprehend speech and read well, but have significant difficulty in repeating phrases.
\end{itemize}

\section{Choosing Transcript Length from TED Talks}
\label{app:length}
We compare differences in values of the 8 POS speech features from speech samples, for transcript lengths of 5, 25, 50, 75 and 100 utterances each. We compute $t$-tests between features computed from transcripts lengths of 5 and 25, 25 and 50, 50 and 75, and 75 and 100. We find that while features 5 out of 8 features are significantly different between transcript lengths of 5 and 25, they stabilize for lengths greater than or equal to 25, i.e, no significant difference between lengths of 25 and 50 (lowest p-value is 0.22), 50 and 75 (lowest p-value is 0.32) and 75 and 100 (lowest p-value is 0.59). Thus, we choose 25 utterances to be the standard length of a transcript from the TED Talks dataset to maximize data available.

\section{Part-of-Speech Proportions Comparisons Across Languages}
\begin{table}[h!]
\begin{adjustbox}{max width=0.75\textwidth}
\scriptsize{\caption{Significant p-values corresponding to T-tests of the 8 features between English and other languages (after Bonferroni correction). Indicated by `*' if significantly different for both Mandarin and French, `+' if only significantly different between English and Mandarin, `\#' if only significantly different between English and French and `-' if there is no significant difference.\label{tab:pvalues}
}}
\begin{tabular}[ht!]{lccc}
\hline
POS/Feature  &  Aphasia & Control\\
\hline
\hline
Nouns & + & +\\
Verbs & * & *\\
Subordinating conjunctions & * & *\\
Adjectives & + & *\\
Adverbs & * & * \\
Co-ordinating Conjunctions &+ & *\\
Determiners & + & *\\
Pronouns &+ & +\\
\end{tabular}
\end{adjustbox}
\end{table}

\section{Hyperparameters}
\label{app:ot_hyperparams}
For EMD, method proposed by \cite{ferradans2014regularized} is used for out of sample mapping to apply to transport samples from a domain into the other with other default parameters in \cite{flamary2017pot}. 

For EMD-R, entropic regularization parameter is set to 3 with all other parameters default.

For Gaussian mapping, the weight for linear OT loss is set to 1, and maximum iterations is set to 20, with stop threshold for iterations set to $1e-05$ with other default parameters in \cite{flamary2017pot}.

\section{Paired Data Does Not Improve Performance Significantly}
We observe, from Tab.~\ref{tab:paired_ot}, that paired data does not significantly improve performance for French or Mandarin, over classification with unpaired datasets for OT.
\begin{table*}[ht]
\caption{F1 macro scores across languages with OT, with paired data.\label{tab:paired_ot}}
{
\begin{adjustbox}{max width=0.7\textwidth}
\begin{tabular}{c|l|c|c|c}
\hline
  \multicolumn{1}{l}{\textbf{Language}} &
  \multicolumn{1}{c}{\textbf{Method}} &
  \multicolumn{1}{c}{\textbf{SVM}}& 
  \multicolumn{1}{c}{\textbf{RF}}&
  \multicolumn{1}{c}{\textbf{NN}} \\ 
  \hline
& & F1 &F1  &F1 \\ 
  
  \hline \hline
 
  \multirow{2}{*}{\makecell{French}} 
  & $OT$-EMD & $\mathbf{83.22\pm 0.00}$ & $ \mathbf{84.71 \pm 1.95}$& $67.22\pm 2.65$ \\
  & $OT$-Gaussian & $46.67\pm 0.00$ & $41.89 \pm 3.38$& $34.12\pm 3.80$ \\
  & $OT$-EMD-R & $\mathbf{83.22\pm 0.00}$ & $ 78.84\pm 0.00$& $ 74.97 \pm 3.41$ \\
  \hline
   \multirow{2}{*}{\makecell{Mandarin}} 
  & $OT$-EMD & $ 59.28\pm 0.00$ & $ \mathbf{55.35\pm 3.38}$& $49.30\pm 2.42$ \\
  & $OT$-Gaussian & $25.97\pm 0.00$ & $25.53 \pm 0.63$& $24.64 \pm 0.00$ \\
  & $OT$-EMD-R & $\mathbf{60.11 \pm 0.00}$ & $ 49.25 \pm 1.26$& $\mathbf{56.12 \pm 2.01}$\\
\end{tabular}
\end{adjustbox}
}
\end{table*}

\section{Studying Effect of Data on Unilingual Performance}
\label{app:data_impact_unilingual}

\begin{figure}[h]
\centering
    \includegraphics[width=3.0in]{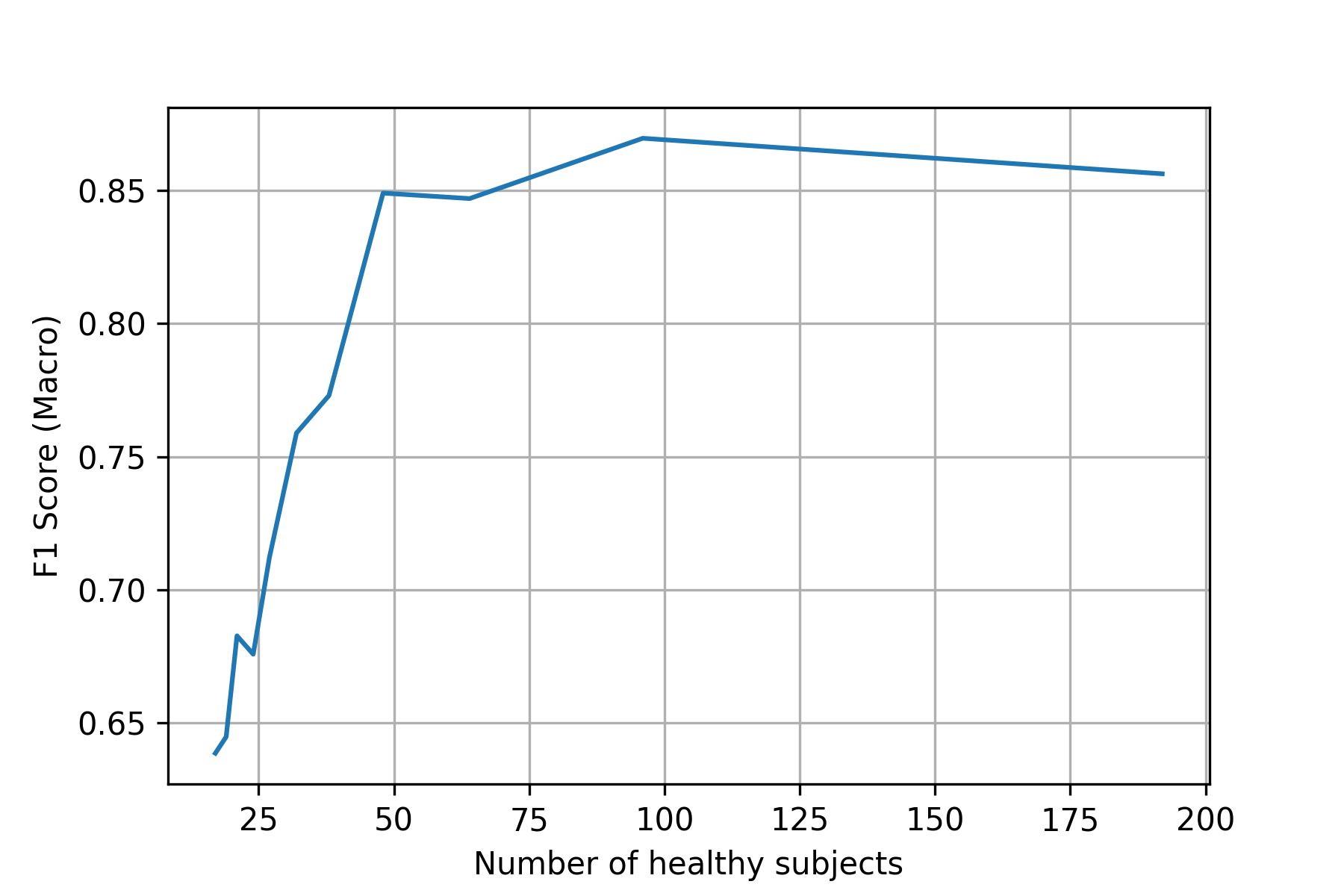}
    \caption{Effect of dataset size on the aphasia detection task.\label{fig:f1_moving}}
\end{figure} 
To study the impact of data on the aphasia detection task, we perform an ablation study wherein the size of the English AphasiaBank dataset is artificially reduced by integer factors (while keeping the relative proportion of healthy and aphasic subjects same). We perform 10-fold cross-validation for a SVM classifier, with progressively less data.  We observe that speech transcripts from atleast 50 healthy subjects are required for the classification performance to stabilize, given the current feature set (see Fig.~\ref{fig:f1_moving}).
F1 scores (micro and macro) increase non-linearly with the addition of data.

\section{Accent Annotation}
\label{app:accent_annot}
The TED-Talks dataset covers a wide speaker demographic, in terms of sex, age and accents. Since prior literature shows that accents can have a significant effect on linguistic features~\cite{runnqvist2013disadvantage}, we manually annotate presence or absence of a North-American accent for English speech in the TED-Talks dataset. An annotator listens to the audio associated with each TED-Talk and annotates if the accent is 'North American' or not. In cases where the accent is not clear, publicly available information regarding nationality of speaker is referenced.
In future work, we plan to have multiple annotations per audio, and factor in metrics such as cross-rater agreement into our analysis.
\end{document}